\documentclass[twocolumn,pra,showpacs,groupaddress]{revtex4}
\pdfoutput=1
\usepackage{amssymb}
\usepackage{amsmath}
\usepackage{graphicx}
\usepackage{bbold}
\usepackage{amsfonts}
\usepackage{mathrsfs}
\usepackage{color}

\begin{document}

\title{Nematic Ferromagnetism on the Lieb lattice}

\author{CHEN Ke-Ji}
\author{ZHANG Wei}
\email{wzhangl@ruc.edu.cn}
\affiliation{Department of Physics, Renmin University of China, Beijing 100872,
People's Republic of China}
\affiliation{Beijing Key Laboratory of Opto-electronic Functional Materials and Micro-nano Devices, 
Renmin University of China, Beijing 100872, People's Republic of China}


\begin{abstract}
We discuss the properties of ferromagnetic orders on the Lieb lattice
and show that a symmetry protected quadratic-flat band crossing point
will dramatically affect the magnetic ordering.
In the presence of a weak on-site repulsive interaction, the ground state 
is a nematic ferromagnetic order with simultaneous
broken of time-reversal and rotational symmetries.
When the interaction strength increases, the rotational symmetry will restore
at a critical value, and the system enters a conventional ferromagnetic regime.
The mean-field transition temperatures for both the nematic and conventional 
ferromagnetic phases are in the order of interaction.
This observation suggests that these magnetic orders have the potential to be realized
and detected in cold atomic systems within realistic experimental conditions.
\end{abstract}

\maketitle

The investigation on ferromagnetism is one of the central topics
in condensed matter physics, and has attracted attention for nearly
a century since the early age of quantum theory.
Heisenberg showed that a system of localized spins would favor
a fully polarized state by gaining exchange energy.~\cite{heisenberg-28}
However, the same argument can not be simply applied in itinerant fermions,
where the kinetic energy of the underlying system
has to be considered on an equal footing with the interaction effect.

The discussion on the stability of itinerant ferromagnetism in a lattice system
dates back to 1960's. Thouless~\cite{thouless-65} and Nagaoka~\cite{nagaoka-66}
pointed out that a ferromagnetic ground state can be stabilized
in any finite bipartite lattice with an infinite on-site repulsive interaction.
Lieb showed that the stable region of ferromagnetism can be extended
to arbitrary repulsive interaction, provided that the number of sites
are different for the composite sublattices.~\cite{lieb-89}
The key ingredient in Lieb's argument is the existence of a non-dispersive,
or equivalently flat, band. When the flat band is partially filled, fermions tend to
be spin polarized to minimize the interaction energy, without paying any cost
in the kinetic energy. In other words, since states in the flat band consist
of localized Wannier functions, the ferromagnetic states can benefit from
the exchange interaction as pointed out by Heisenberg.
Subsequent studies confirm the stability of ferromagnetism in
various models~\cite{mielke-91a,mielke-91b,mielke-92, tasaki-92, tasaki-95, tasaki-07}, and generalize the idea
to nearly-flat-band cases~\cite{tasaki-96}.
Experimental realization of the (nearly) flat-band ferromagnetism
has been proposed in a class of physical systems including
atomic quantum wires~\cite{arita-98},
quantum-dot super-lattices~\cite{tamura-02},
and organic polymers~\cite{suwa-03}.

\begin{figure}[tbp]
\begin{center}
\includegraphics[width=8cm]{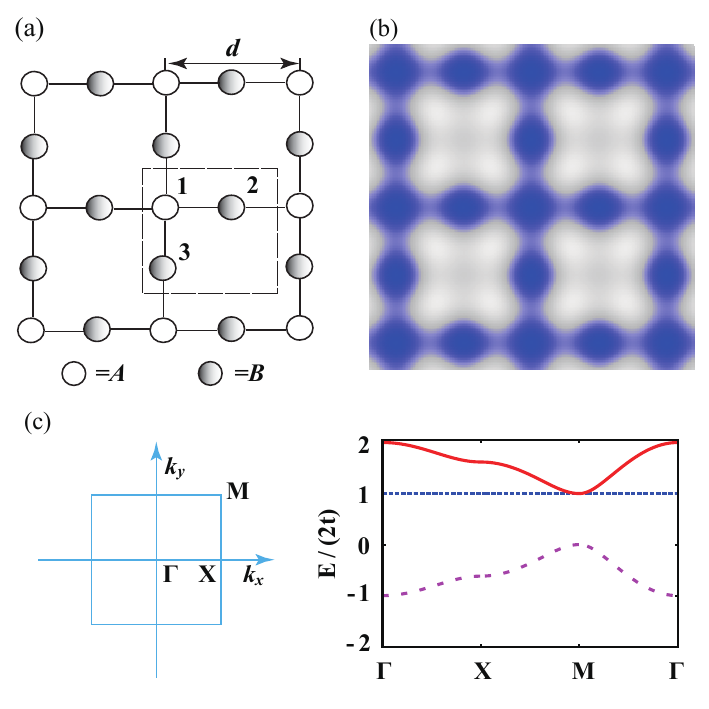}\\
\caption{(Color online) (a) 2D Lieb lattice consisting of two square
sublattices A and B. There are three sites (1, 2, 3) within a unit
cell (dotted square).
(b) The Lieb lattice can be realized by arranging
three square optical lattices. Here, a typical example with
$V_{10} = V_{20} = 2 V_{30}$ (see text) is shown to demonstrate the
case of $V_b/(2t) = 0$.
(c) Band structure of the Lieb lattice with $V_b/(2t) = 1$. Note that the
flat band degenerates with the upper dispersive band at
the ${\bf M}$-point, leading to a quadratic-flat band crossing point.}
\label{fig-lattice}
\end{center}
\end{figure}

Rapid progress in cold atom experiments has paved a new route
towards the exploration of ferromagnetism in itinerant fermions.
Thank for the extraordinary controllability of lattice potentials and interaction,
several proposals have been given to realize (nearly) flat-band
ferromagnetism~\cite{wang-08, noda-09, zhang-10}.
In particular, Noda {\it et al.} investigated two-component cold
fermions loaded into a two-dimensional (2D) generalized Lieb lattice,
and suggested that a ferromagnetic order can be stabilized in
a wide parameter region.~\cite{noda-09}
The 2D generalized Lieb lattice consists of two square lattices
(sublattice A and B), and has three sites per unit cell, as depicted
in Fig.~\ref{fig-lattice}(a). With only nearest-neighbor hopping,
the lattice topology supports a flat band in the middle of two
dispersive bands, hence can stabilize ferromagnetism when it is
partially filled.

Apparently, a band structure containing only one single flat band does not exist in any realistic 
physical system. In all proposed model lattices for flat-band ferromagnetism,
the flat band is always associated with dispersive bands,
and the effect of their accompany is not fully understood.
As a typical example, if the two composite sublattices of the
Lieb lattice have the same depth, the flat band degenerates with the
two linearly dispersive bands at the ${\bf M} = (\pi/d,\pi/d)$ point.
On the other hand, in general cases where the depth of
sublattice B is shifted from that of sublattice A by an amount of $V_b$,
the flat band touches only one of the two dispersive bands
at the ${\bf M}$-point, as illustrated in Fig.~\ref{fig-lattice}(c),
leading to a quadratic-flat band crossing point (QFBCP).

In this manuscript, we show that the ferromagnetic order is
dramatically affected by the existence of the QFBCP.
In the non-interacting level, the QFBCP is protected
by the time reversal (TR) and $C_4$ rotational symmetries.
When a repulsive on-site interaction is present, since
the density of state of the flat band is singular, the QFBCP becomes
marginally unstable, leading to a spontaneous broken of
the TR and/or $C_4$ rotational symmetries.
At half filling, we find that in the weak coupling limit
the ground state is a nematic ferromagnetic (NFM) order,
where the spontaneous magnetization on sites 2 and 3 are different such that
the $C_4$ rotational symmetry of the underlying lattice
is broken down to $C_2$.
The $C_4$ rotational symmetry will restore with increasing interaction
via a second or first order phase transition, depending on the
value of $V_b$. Within a mean-field calculation,
we further map out the phase diagram,
and identify three phases including: (i) a semimetal with NFM order;
(ii) a band insulator with NFM order, and (iii) a band insulator
with conventional FM order. These magnetic orders have the potential
to be realized and detected in cold fermions loaded in optical lattices.

We consider spin $1/2$ fermions loaded in a 2D Lieb lattice
\begin{eqnarray}\label{eqn:Hamiltonian}
H &=& -t \sum_{<i,j>,\alpha} c_{i \alpha}^\dagger c_{j \alpha}+ V_b \sum_{i \in B} n_i \nonumber\\
&& + U_{A} \sum_{i \in A} n_{i \uparrow} n_{i \downarrow}
+U_{B}\sum_{i \in B}n_{i \uparrow}n_{i \downarrow},
\end{eqnarray}
where $t$ is the nearest neighbor hopping matrix, $c_{i \alpha}^\dagger (c_{i \alpha})$
is the creation (annihilation) operator for fermion on site $i$ with spin $\alpha$,
$V_b\equiv \epsilon_{B}-\epsilon_{A}$ is the chemical potential offset for the sublattice B
(i.e., the relative shift of the two sublattices), $\epsilon_{i=A,B}$ describes an energy of each lattice site,
and $U_{i=A,B}$ is the on-site interaction. In the context of cold atoms,
a Lieb lattice with lattice parameter $d$ can be realized by
arranging three square optical lattices with
$V_1 = V_{10} [\sin^2(2 \pi x/d) + \sin^2(2 \pi y/d)]$,
$V_2 = V_{20} [\sin^2(\pi x/d + \pi) + \sin^2(\pi y/d + \pi)]$, and
$V_3 = V_{30} [\cos^2(\pi x/d + \pi y/d) + \cos^2(\pi x/d - \pi y/d)]$.
In this configuration, the potential depths for sublattices A and B are
$h_A = V_{10} + V_{20}/2 - V_{30}$ and $h_B = V_{10} - V_{20}/2 + V_{30}$,
respectively, and can be controlled independently via suitable combination
of laser intensities~\cite{goldman-11}. For an optical lattice given by the above configuration, 
parameters of Eq. \eqref{eqn:Hamiltonian} can be attained by using harmonic approximation
where the Wannier function on a lattice site is approximated by the ground state wavefunction
of a harmonic oscillator~\cite{jaksch-98}.
Under this approximation, the on-site interaction can be rewritten as 
$U_{i}/ E_{\bf r}=(2h_{i}/\pi E_{\bf r})^{1/2}a_s/a_z$,
where the recoil energy $E_{\bf r} \equiv \pi^{2}\hbar^{2}/(2ma^{2})$ is used as the 
energy unit and $a_z$ is the characteristic length of $z$-axis trapping potential. 
Similarly, $t$ and $V_{b}$ can also be obtained analytically. 
According to the calculation above, this arrangement offers the ability to tune the hopping
matrix $t$ and the potential offset $V_b$.
A repulsive interaction $U_{i}>0$
can be achieved and varied via an adiabatic ramping to the upper branch
on the BEC side of a Feshbach resonance~\cite{jo-09}.

In the non-interacting case, the Hamiltonian can be diagonalized in
momentum space, leading to a band structure consisting of three bands
as shown in Fig.~\ref{fig-lattice}(c). One of the three bands is completely flat,
as required by the bipartiteness.
The flat band has Bloch wavefunction
$\propto [0, -\cos(k_x a), \cos(k_y a)]$ on the three sites within a unit cell,
indicating the presence of local Wannier functions residing on sublattice B
and having opposite amplitudes between sites 2 and 3.
When the chemical potential offset $V_b = 0$, the three bands
are degenerate at the ${\bf M} = (\pi/d, \pi/d)$ point, where the two dispersive
bands linearly intersect with the flat band. In general cases of $V_b \neq 0$,
depending on the sign of $V_b$, one of the two massive bands breaks the
${\bf M}$-point degeneracy, and the flat band only touches the other dispersive band
as illustrated in Fig.~\ref{fig-lattice}(c). As a consequence, the ${\bf M}$-point
becomes a quadratic-flat band crossing point (QFBCP).
Around this point, the effective two-band Hamiltonian reads
\begin{equation}
\label{eqn:effH}
H_0^{\rm eff} = \frac{2 t^2}{V_b} \left(\begin{array}{cc} |\delta {\bf k} |^2 & \delta {\bf k}_+^2 \\
\delta {\bf k}_-^2 & |\delta {\bf k} |^2 \end{array} \right) + {\cal O}(\delta {\bf k}^4),
\end{equation}
where $\delta_{\bf k} = {\bf k} - {\bf M}$ and
$\delta {\bf k}_\pm = \delta k_x \pm i \delta k_{y}$.
From this effective Hamiltonian, it is clear that the cases of positive and
negative chemical potential offset $V_b$ are equivalent via a particle-hole transformation.
Thus, we focus on systems with $V_b >0$ without loss of generality
in the following discussion.

The presence of a QFBCP at the ${\bf M}$-point is the central feature of the Lieb lattice.
This band crossing point (BCP) is protected by the TR and $C_4$
rotational symmetries in the non-interacting case, and is characterized with a
nontrivial topological index $2\pi$.
Such a putative topologically stable BCP becomes marginally unstable against
infinitesimal repulsive interaction~\cite{sun-09},
leading to a spontaneous broken of TR and/or $C_4$ rotational symmetries,
which drives the system towards a magnetic and/or nematic phase.
Besides, since the BCP consists of a
non-dispersive band, the infinite density of state allows the possibility
of filling the higher momentum states with one single spin species without gaining
any kinetic energy. As a consequence, if the on-site interaction $U_{i}$ is repulsive,
the system could easily favor a ferromagnetic phase
for filling factors between 1/3 and 2/3.
\begin{figure}[tbp]
\begin{center}
\includegraphics[width=8cm]{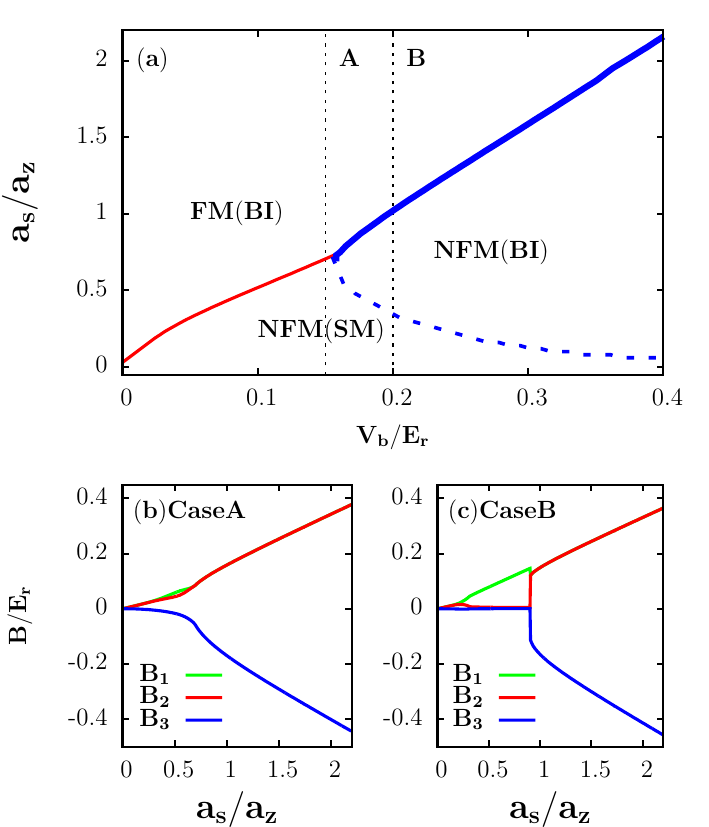}\\
\caption{(Color online) (a) Zero temperature phase diagram for spin $1/2$ Fermi system
on the Lieb lattice at half filling. The onsite magnetization within sublattice B
preserves the lattice symmetry within the FM phase, and becomes
non-uniform in the NFM regime. The NFM phase is favored for
weak interaction, and is separated from the FM regime
via a second order (thin solid) or first order (thick solid) transition.
Within the NFM regime, the system is a semimetal (SM) for small $a_s/a_z$ and $V_b/E_{\bf r}$,
and becomes a band insulator (BI) with increasing $a_{s}/a_{z}$ and $V_b/E_{\bf r}$.
For two typical values of $V_b/E_{\bf r} = 0.15$ and $0.2$,
the magnetization on the three sites are shown in (b) and (c), respectively. Here, we fixed $V_{10}=0.5E_{\bf r}$ and $V_{30}=0.1E_{\bf r}$.
}
\label{fig-phase}
\end{center}
\end{figure}

Next, we focus on the case of repulsive Hubbard $U_{i}>0$ at half filling,
and investigate the possibility for a stable ferromagnetic order.
Within a mean-field (MF) level, the magnetic order is characterized
by on-site magnetization ${\bf B}_{i=1,2,3} \equiv - 2 U_{i} \langle {\bf S}_i \rangle/(3\hbar)$,
where ${\bf S}_i$ is the spin operator on site $i$.
The momentum-space Hamiltonian takes the form
\begin{equation}
H_{\bf k}^{\rm MF} = \left(
 \begin{array}{ccc}
{\bf B}_1 \cdot \vec{\sigma}  & -2t\cos{k_x} \cdot \mathbb{1} & -2t\cos{k_y} \cdot \mathbb{1} \\
 & {\bf B}_2 \cdot \vec{\sigma} + V_b \cdot \mathbb{1} & 0 \\
 &  & {\bf B}_3 \cdot \vec{\sigma}  + V_b \cdot \mathbb{1}
 \end{array}
\right),
\end{equation}
where $\vec{\sigma} = (\sigma_x, \sigma_y, \sigma_z)$ represents Pauli matrices,
$\mathbb{1}$ denotes identity in the spin space, and the lower off-diagonal
elements are filled to guarantee the Hermiticity of the Hamiltonian.
The six energy bands of this Hamiltonian are occupied to half filling, leading to
the energy functional
\begin{eqnarray}
\label{eqn:energy}
{\mathcal E} &=& \sum_{({\bf k}, j) \in \Gamma} E_{j} ({\bf k})
+ \frac{N_{0}}{2U_{A}}B^{2}_{1}+\frac{N_{0}}{2U_{B}}(B^{2}_{2}+B^{2}_{3})\nonumber\\
&&+\frac{1}{6}N_{0}U_{A}+\frac{1}{3}N_{0}U_{B},
\end{eqnarray}
where the summation over $({\bf k}, j) $ is restricted
to the set of occupied Bloch states $\Gamma$ with band index $j$,
and $N_0$ denotes the total number of unit cells.

To study the possibility of nematic magnetic order with both TR and $C_4$ rotational
symmetries broken, we allow the magnetization on sublattice B can be different for
sites 2 and 3, and map out the zero-temperature phase diagram
as shown in Fig.~\ref{fig-phase}(a) by minimizing the energy functional.
We find that the optimized magnetization ${\bf B}_{i=1,2,3}$ are always
along the $z$-axis, hence we consider only axial magnetic order in the
following discussion.

In the weakly interacting limit, the nematic ferromagnetic phase (NFM)
is always favorable with an exponentially small magnetization difference.
When $a_{s}/a_{z}$ is increasing from the weakly interacting limit,
the NFM phase remains stable up to a critical value $(a_{s}/a_{z})_{c}$,
above which the $C_4$ rotational symmetry restores
and the system enters the FM regime.
The order of the NFM-FM phase transition depends
on the value of $V_b$. For $V_b$ below a critical value $V_c/E_{\bf r} \approx 0.156$,
the NFM-FM transition is of the second order,
as identified by the condition $V_b = B_{1z} - B_{2z}$ and
depicted by a thin solid line in Fig.~\ref{fig-phase}(a).
By increasing $V_b > V_c$, the NFM-FM phase boundary becomes
of the first order [thick solid line Fig.~\ref{fig-phase}(a)],
resulting from the competition
between the corresponding metastable states.

\begin{figure}[tbp]
\begin{center}
\includegraphics[width=8cm]{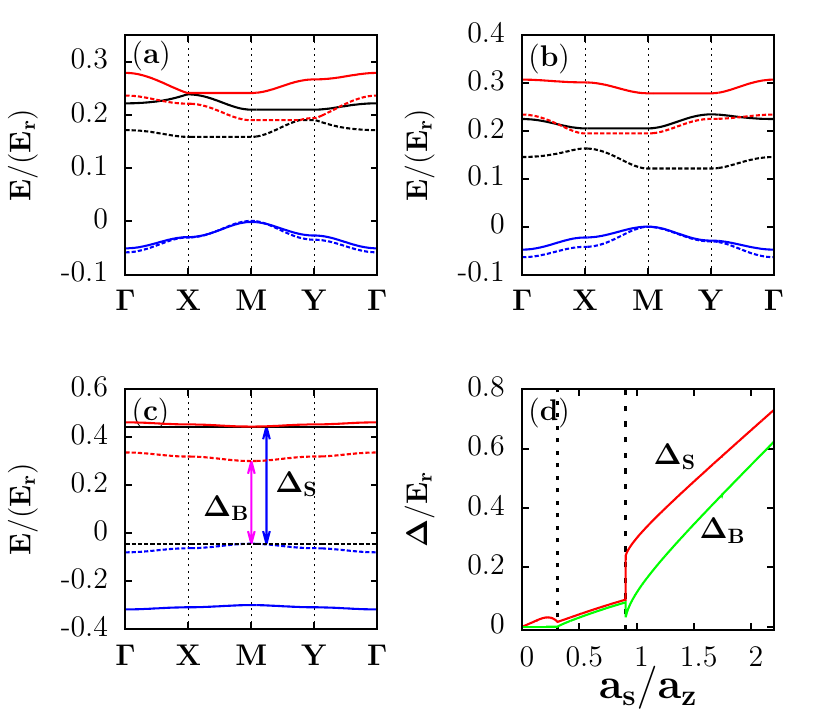}
\caption{(Color online) Band structure for (a) semimetal with NFM order,
(b) band insulator with NFM order,
and (c) band insulator with FM order.
The solid (dashed) curves represent up (down) spin.
These three cases are represented in the phase diagram Fig.~\protect\ref{fig-phase}(a), 
and correspond to $V_b/E_{\bf r} = 0.2$, $a_{s}/a_{z} = 0.3$, $0.5$ and $1.5$, respectively.
The variation of $\Delta_s$ and $\Delta_b$ for $V_b/E_{\bf r} = 0.2$
is shown in (d). Note that $\Delta_s$ is always in the order of the interaction. 
Here, $V_{10}=0.5E_{\bf r}$ and $V_{30}=0.1E_{\bf r}$ are fixed. }
\label{fig-band}
\end{center}
\end{figure}

Within the NFM phase, the $C_4$ rotational symmetry
is spontaneous broken down to $C_2$ by splitting the QFBCP into
two Dirac points located along the direction of one of the principal axes
for weak interactions. The two Dirac points have the same Berry flux $\pi$,
in clear contrast to the case of graphene where the two Dirac points have
Berry fluxes $\pi$ and $-\pi$. In this case, the system is an anisotropic
semimetal (SM) at half filling, with the Fermi surface shrinks to the two Dirac
points as shown in Fig.~\ref{fig-band}(a). By increasing $a_{s}/a_{z}$,
the two Dirac points move towards the boundary of the Brillouin zone,
and eventually disappear when the magnetization difference
$\delta = | B_{2z} - B_{3z} |$ between sites 2 and 3
exceeds the bandwidth of the first excited dispersive band. As a result, a full gap
is open and the system becomes an anisotropic band insulator (BI) as
depicted in Fig.~\ref{fig-band}(b). When the system enters the FM regime,
the magnetization is large enough such that a finite gap is always present,
and the system is a band insulator as shown in Fig.~\ref{fig-band}(c).
As the system enters the band insulator regime, being either a NFM or a FM state, 
we expect a quantum anomalous hall effect state can be stabilized provided 
a weak spin-orbit coupling~\cite{wu-11} that supports a non-trivial Chern number of the occupied band
but does not close the band gap, as discussed in Ref.~\cite{zhao-12}.

Note that in both the NFM and FM phases, the spin gap
to create a minority spin $\Delta_s$ remains in the order
of the interaction strength $U$ [See Fig.~\ref{fig-band}(d)].
Although the Mermin-Wigner theorem excludes the possibility of
any 2D ferromagnetic order at finite temperature in the thermodynamic limit,
the existence of such an order in a finite size system is perfectly allowed,
provided that the coherence length $\xi$ is comparable or exceeding the
system size $L$. Specifically, the coherence length of the magnetic order is
$\xi \sim \hbar c \exp(\rho_s/T)/(k_B T)$, where $c$ is the spin-wave
velocity at zero temperature, and the phase stiffness $\rho_s$
is in the same order of the MF transition temperature $T_{\rm MF}$~\cite{chakravarty-88}.
Therefore, when the temperature goes well below $\Delta_s$, the coherence
length can greatly exceed the lattice spacing such that domains with
magnetic ordering can be formed. Since the energy gap $\Delta_s$ increases
almost linearly with $U$, and the interaction can be tuned to be fairly
large via an $s$-wave Feshbach resonance, the temperature required
to observe the FM and NFM domains could be reachable within present
technology.

In a realistic setup of cold atom experiment, a 2D geometry proposed in our
manuscript is usually realized by a one-dimensional optical lattice or
a pancake-shaped trap. The resulting trapping potential is highly anisotropic
with typical trap size $L_z \sim \mu$m along the strongly confined axial direction
and $L_{xy} \sim 10^2$ $\mu$m in the quasi-2D radial plane. To incorporate
the global trapping potential, one commonly used scheme is to implement the
local density approximation (LDA) by introducing a position dependent chemical
potential $\mu({\bf r})$. The LDA is valid as the coherence length is
much smaller than the radial trap size $\xi \ll L_{xy}$.~\cite{lda}
Besides, we also require the coherence length to be much greater than the
lattice spacing $\xi \gg d$, such that FM or NFM ordered domains can be formed
and resolved. Since the lattice spacing $d \sim 10^2$ nm is about 3 orders of
magnitude smaller than $L_{xy}$, there is a large parameter window for the system
to establish magnetic orders within LDA, where different phases coexist and
form a ring structure.

The detection of the ferromagnetic orders can be implemented
via an {\it in-situ} measurement~\cite{chin-09, greiner-10},
which is able to extract single site density distribution for different spin species,
and hence the local magnetization
${B}_i \propto n_{i\uparrow} - n_{i\downarrow}$.
If the system is prepared with equally populated two-component Fermi gas,
we expect to resolve FM or NFM domains with opposite magnetizations.
Another possible detection scheme is to measure the single-particle
dispersion with Bragg spectroscopy~\cite{ernst-10} or
angle-resolved photoemission spectroscopy (ARPES)~\cite{jin-08}.

In summary, we discuss the effect of a quadratic-flat band crossing
point (QFBCP) on the ferromagnetic (FM) order. Taking the
2D Lieb lattice as an example, we show that the QFBCP
is marginally unstable against infinitesimal repulsive interaction,
given an infinite density of state of the flat band.
In the weakly interacting limit, the ground state is a nematic
ferromagnetic (NFM) order with time-reversal and
rotational symmetries broken. Within the NFM regime,
the spontaneous generated magnetizations are different
on sublattice B, and the QFBCP is broken
into two Dirac points along one of the principal axes.
In the strong coupling limit, the interaction $U$ becomes
the only relevant energy scale, and a conventional
ferromagnetic (FM) phase is favored.
We then map out the zero-temperature phase diagram within
a mean-field analysis, and identify three regions including
a semimetal with NFM order, a band insulator with NFM order,
and a band insulator with FM order.
We point out that the spin gap for all three phases is in the same
order of interaction, which can be tuned via
a Feshbach resonance. Thus, we expect these magnetic phases
can be realized in two-component Fermi gases loaded in optical
lattices at experimentally reachable temperatures, and can be
distinguished via a species selective {\it in-situ} measurement.

This work is supported by NSFC (10904172, 11274009), 
the Research Funds of Renmin University of China (14XNH061, 10XNL016),
and the Program of State Key Laboratory of Quantum Optics and 
Quantum Optics Devices (KF201404).

\end{document}